# NewsDeps:
# Visualizing the Origin of Information in News Articles

Felix Hamborg[1], Philipp Meschenmoser[1], Moritz Schubotz[2], Bela Gipp[2]

[1] University of Konstanz, Germany
`{first.last@uni-konstanz.de}`

[2] University of Wuppertal, Germany
`{last}@uni-wuppertal.de`

**Abstract.** In scientific publications, citations allow readers to assess the authenticity of the presented information and verify it in the original context. News articles, however, do not contain citations and only rarely refer readers to further sources. Readers often cannot assess the authenticity of the presented information as its origin is unclear. We present NewsDeps, the first approach that analyzes and visualizes where information in news articles stems from. NewsDeps employs methods from natural language processing and plagiarism detection to measure article similarity. We devise a temporal-force-directed graph that places articles as nodes chronologically. The graph connects articles by edges varying in width depending on the articles' similarity. We demonstrate our approach in a case study with two real-world scenarios. We find that NewsDeps increases efficiency and transparency in news consumption by revealing which previously published articles are the primary sources of each given article.

**Keywords:** News dependency, content borrowing, news plagiarism detection.

## 1    Introduction and Related Work

The rise of online news publishing and consumption has made information from various sources and even other countries easily accessible [5], but has also led to a decrease of reporting quality [10, 13]. The increasing pace of the publish-consume cycle leaves publishers with less time for journalistic investigation and information verification. At the same time, the pressure to publish a story soon after the event has happened rises, as competing outlets will do so likewise. Also, journalists routinely copy-edit or reuse information from previously published articles, which increases the chance of spreading incorrect or unverified information even more [4]. In 2010, a study showed that over 80% of articles reporting on the same topic did not add any new information, but merely reused information contained in articles published previously by other outlets [16].

Currently, regular news consumers cannot effectively assess the authenticity of information conveyed in articles. Understanding where information stems from, and how the information differs from the used sources could help readers to assess the authen-ticity of such information and ease further verification. For the same reasons, documentation of the origin of information is a fundamental standard in scientific writing. News, however, do not contain citations and only rarely refer readers to further articles [1].

The objective of our research is to identify and visualize information reuse (or content borrowing) in articles. We call this task *news information reuse detection* (NIRD). By showing how related articles reuse information, NIRD helps users to assess the authenticity. Additionally, NIRD increases the efficiency of news consumption as users can see if an article contains novel information or is a mere copy of other articles.

Only few NIRD approaches have been proposed, but news dependencies are still non-transparent to users because none of the approaches visualizes the results. One approach employs methods from *plagiarism detection* (PD) to find unauthorized instances of information reuse in articles written in Korean [12]. The approach computes the similarity of each article pair in a set of related articles. A high similarity between two articles suggests that the latter article contains information from the former article. To assess the similarity of two articles, the approach uses *fingerprinting*, a commonly used text similarity measure in PD that identifies n-grams shared by two documents [17]. A similar approach is *qSign*, which finds instances of information reuse in news blogs and articles using also fingerprinting [9]. However, since none of the approaches visualizes dependencies, the origin of information remains unclear in regular news consumption.

In Section 2, we propose *NewsDeps*, a NIRD approach that integrates analysis and visualization of news dependencies. Our main contribution is a visualization that reveals which articles reuse information from other articles. We demonstrate this functionality in a case study in Section 3. Our study shows that NewsDeps also provides an overview of current topics, a common use case in regular news consumption.

## 2 System Overview

NewsDeps is a NIRD approach that aims to reveal information reuse on article level, i.e., identify and visualize the *main sources* of information for each of the articles. When applied to a set of related articles, NewsDeps shows which articles use information from previous articles, and which articles are unrelated. The workflow consists of three phases: (1) *news import*, (2) *similarity measurement*, (3) *visualization*. We describe the second and third phase in the following subsections.

NewsDeps can import articles either from *JSON files* or by *URL import*. The system accepts any document that contains a title, main text, publisher, and publishing date and time. Additional fields, such as to define the background color of each article in the visualization, can also be processed. The URL import allows users to import a list of URLs referring to online articles. The systems then retrieve the required fields using *news-please*, a web crawler and information extractor tailored for news [7].

## 2.1  Measurement of Information Reuse in Articles

Upon user request, NewsDeps computes a $k \times k$ *document-to-document similarity (d2d) matrix*, where $k$ is the number of imported articles. Specifically, we compute only the upper right diagonal half of the d2d matrix, as information can only flow from previous articles to articles published afterward. NewsDeps currently supports two basic text similarity measures (1) *TF-IDF & cosine* and (2) *Jaccard,* and two plagiarism scores provided by the well-established approaches (3) *Sherlock* [8] and (4) *JPlag* [11], which commonly serve as base-line similarity measures in PD. The user can choose which similarity measure to apply. Sherlock and JPlag try to estimate how likely one of two documents has plagiarized from the other (first case in Section 3), while the text similarity measures TF-IDF & cosine and Jaccard are more suitable to group similar articles, e.g., by topic (second case). We linearly normalize all similarity values between 0 and 1, where 1 indicates high similarity between two articles.

## 2.2  Visualization of News Dependencies

NewsDeps visualizes the dependencies of the imported articles in a directed graph (see Fig. 1). The articles are ordered temporally on one axis (by default the x-axis, but the user can swap the function of both axes). Each node represents a single article. We use the width of the edge between two articles as a visual variable to encode the pair's similarity, i.e., the more similar, the thicker the edge. Pairs below a user-defined *similarity threshold* are not connected visually.

The other axis, by default the y-axis, allows placing the article in a way that avoids occlusion. Technically, the graph is created by first placing articles at their temporal position. Afterwards, we apply force-direct node placement [2] in the other dimension, i.e., the articles are moved along the second axis until all articles are placed in a way that their distances to each other best represent their similarities. We call the resulting graph *temporal-force-directed (TFD) graph*. The primary purpose of the TFD graph is to reduce overlap of articles and edge crossings while adhering to chronological order. Note that most visualizations use force-directed graphs to show (dis-)similarity of elements, but the TFD graph's temporal placement of nodes may skew perceived distances between nodes. Therefore, we encode article similarity using the width of the edges.

NewsDeps follows the *Information Seeking Mantra* [15]: overview first, zoom and filter, then details on demand. Therefore, the approach offers four *levels of detail* (LOD) to display articles: *none* (the graph shows each article as a point only), *source* (the node displays an article's publisher), *title* (title only), *detailed* (title, publisher, main picture, and lead paragraph). An *automated* mode automatically chooses proper LOD, by balancing between showing more details and reducing overlap of visuals. NewsDeps provides various interaction techniques [15] to support the visual exploration by the user. For example, both axes and the graph can be dragged and zoomed to adjust the position and granularity of the time frame, details on demand are shown by hovering over important elements, such as articles and edges, and clicking on an article opens a popup showing the full article (see Fig. 2). Each article has a linked axis-indication on each of the axes, which allows users to easily find the article that was published at a specific time (by hovering the article, the indications on both aces are highlighted), and vice versa. Users can set the opacity of nodes, edges, and axis-indications, which helps with occlusion that can occur in analyses of many articles. The edges are naturally directed as to the chronological order as articles can only reuse information from previous articles.

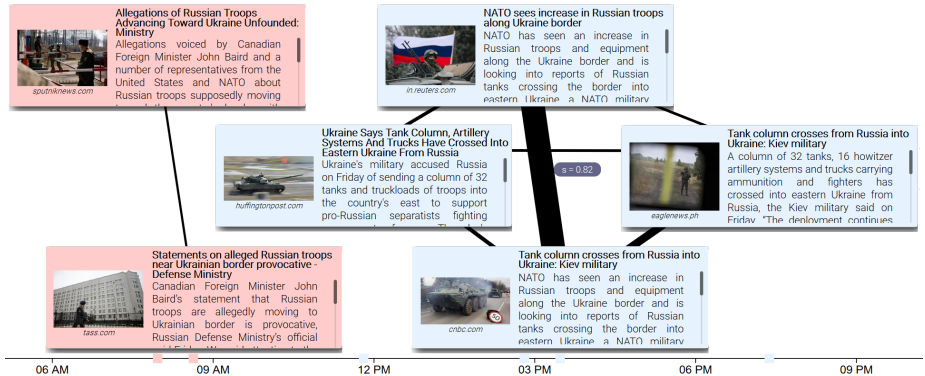

**Fig. 1.** NewsDeps shows articles as nodes in a temporal-force-directed graph.

## 3 Case Study

We analyze the strengths and weaknesses of our approach in a case study of two real-world news scenarios. In the first case, we investigate how NewsDeps supports exploration of news dependencies. Fig. 1 shows the first case: six articles reporting on the same event during the Ukraine crisis on November 11, 2014. We use JPlag as we are interested in the likelihood that one article reused information from another. The edges in Fig. 1 show that Western outlets (blue articles) reused information from other Western outlets but not or to a lesser extent from Russian outlets (red articles), and vice versa.[1] Hence, Western and Russian articles form separate, mostly not interconnected groups. The thick edge in the middle reveals that the CNBC article contains a large portion of information from the Reuters press release: the authors only changed the title and few other paragraphs towards the end of the article; the lead paragraph is a 1:1 copy, see the labels of both nodes. NewsDeps additionally reveals content differences through the detailed LOD: Western media reported that Russian troops invaded Ukraine, while the Russian state news agency TASS contrarily stated that even the claims that tanks were *near* the borders were false.

In the second case, we investigate how NewsDeps supports regular news consumption. We use TF-IDF & cosine since we are interested in grouping articles reporting on the same topic, and not to find instances of information reuse. Fig. 2 shows three clusters of interconnected articles, each cluster reporting on a separate topic on May 19th, 2017. With the LOD set to title, the visualization helps to get an overview of the different topics quickly. The TFD graph helps to see when coverage on each of the topics began, and when new articles were published. The temporal placement also allows users to track when and what information is added to the coverage. For example, early articles of the blue topic state that a Turkish passenger was detained on a flight, while the last article adds more detailed information on how the passenger was fixated. Finally, the full article popup allows users to read the whole story.

---

[1] This phenomenon has been studied as *media bias by source selection* in the social sciences.

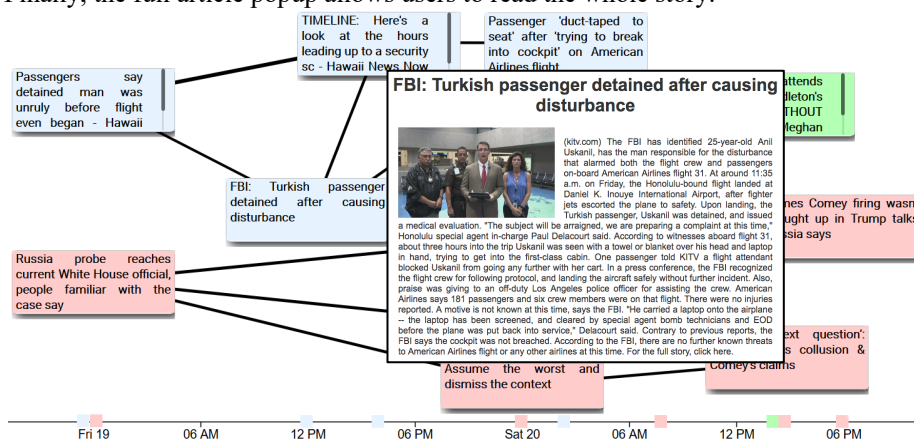

**Fig. 2.** Three groups of articles, each representing a different topic. The popup shows the article.

## 4 Discussion and Future Work

NewsDeps is a first step in the direction of semi-automated NIRD. The first case in our study demonstrated that NewsDeps helps readers to *differentiate* between articles that add new information or present the event from a different perspective and others that merely repeat previously published information. Hence, the visualization *increases the transparency* because users can follow the dissemination and determine the origin of information effectively in a group of related articles. NewsDeps can also be helpful to researchers concerned with relations and dependencies of articles and publishers, such as in the social sciences. The second case showed that NewsDeps also allows getting *an overview* of *multiple topics*, which is the primary purpose of popular news aggregators, such as Google News. However, to quickly get an overview of many articles, the visualization currently lacks cluster labels. By summarizing a topic with one label, readers would also be able to get an overview of large scale news scenarios.

The first case of our case study showed that NewsDeps detects simple forms of information reuse. In the future, we plan to investigate how to identify also more complex forms, such as paraphrased articles or when authors omitted single sentences or paragraphs. We plan to analyze the events described in articles: an event can be described by answers to the five journalistic W and one H-questions (5W1H), i.e., *who did what, when, where, why,* and *how* [6, 14]. Instead of analyzing all tokens in the text, we plan to conceive similarity measures that analyze and compare the 5W of described events in two articles.

NewsDeps currently allows users to analyze dependencies on article level. Following the overview first, details on demand mantra [15], we plan to add a visualization

6where users can select an article and compare it with its main sources. The visualization will then enable the users to view which information has been *committed* or *omitted* compared to the sources. PD visualizations commonly allow for detailed comparison of a selected document and suspicious documents [3], in our case the main sources.

## 5 Conclusion

In this paper, we described NewsDeps, the first information reuse detection system for news, which combines analysis and visualization of news dependencies on article-level. With the help of state-of-the-art similarity measures, our system determines the main sources of information for each of the analyzed articles. The visualization shows all articles in a temporal-force-directed (TFD) graph, which reveals if and how strongly articles relate to another. By showing how related articles reuse information, NewsDeps helps users to assess the authenticity. Additionally, NewsDeps increases the efficiency of news consumption as users can see if an article contains novel information or is a mere copy of other articles. In a case study, we demonstrated that NewsDeps reveals which articles repeat known information, and which articles contain new information or show a different perspective than previously published articles. Hence, the TFD graph helps to understand the dependencies in news coverage on article level.